\documentclass[12pt,a4paper,final]{iopart}

\usepackage{iopams}  
\usepackage{float}  
\usepackage{graphicx}
\usepackage[breaklinks=true,colorlinks=true,linkcolor=blue,urlcolor=blue,citecolor=blue]{hyperref}
\usepackage{color}
\usepackage{url}

\begin{document}

\title[Modeling electron temperature profiles in the pedestal with simple formulas for ETG transport]{Modeling electron temperature profiles in the pedestal with simple formulas for ETG transport}

\author{D. R. Hatch$^{1*,2}$}
\author{M. T. Kotscenreuther$^{2,1}$}
\author{P.-Y. Li$^{1}$}
\author{B. Chapman-Oplopoiou$^{3}$}
\author{J. Parisi$^{4}$}
\author{S. M. Mahajan$^{1,2}$}
\author{R. Groebner$^{5}$}
\address{$^1$Institute for Fusion Studies, University of Texas at Austin, Austin, Texas, USA}
\address{$^2$ExoFusion, Austin, Texas, USA}
\address{$^3$UKAEA-CCFE, Culham Science Centre, Abingdon, OX143DB, UK}
\address{$^4$Princeton Plasma Physics Laboratory, Princeton, New Jersey, USA}
\address{$^5$General Atomics, San Diego, CA, USA}
\ead{$^*$drhatch@austin.utexas.edu}



\begin{abstract}
This paper reports on the refinement (building on Ref.~\cite{hatch_22}) and application of simple formulas for electron heat transport from electron temperature gradient (ETG) driven turbulence in the pedestal.  The formulas are improved by (1) improving the parameterization for certain key parameters and (2) carefully accounting for the impact of geometry and shaping in the underlying gyrokinetic simulation database.  Comparisons with nonlinear gyrokinetic simulations of ETG transport in the MAST pedestal demonstrate the model's applicability to spherical tokamaks in addition to standard aspect ratio tokamaks.  We identify bounds for model applicability: the model is accurate in the steep gradient region, where the ETG turbulence is largely slab-like, but accuracy decreases as the temperature gradient becomes weaker in the pedestal top and the instabilities become increasingly toroidal in nature.  We use the formula to model the electron temperature profile in the pedestal for four experimental scenarios while extensively varying input parameters to represent uncertainties. In all cases, the predicted electron temperature pedestal exhibits extreme sensitivity to separatrix temperature and density, which has implications for core-edge integration.  The model reproduces the electron temperature profile for high $\eta_e = L_{ne}/L_{Te}$ scenarios but not for low $\eta_e$ scenarios in which microtearing modes have been identified.  We develop a proof-of-concept model for MTM transport and explore the relative roles of ETG and MTM in setting the electron temperature profile.  We propose that pedestal scenarios predicted for future devices should be tested for compatibility with ETG transport.             

\end{abstract}

\vspace{2pc}
\section{Introduction} \label{sec:introduction}











Transport in the H-mode pedestal is produced by multiple mechanisms in the various transport channels~\cite{kotschenreuther_19,TPT}.  This paper describes improved algebraic expressions for transport from electron temperature gradient (ETG) driven turbulence, which is often a major transport mechanism in the electron thermal channel.  ETG turbulence has been an active area of research for several decades, starting with initial gyrokinetic demonstrations of its relevance counter to naive scaling estimates~\cite{jenko_00b,dorland_prl_00}.  Recently, ETG studies have focused on the pedestal~\cite{told_08,jenko_09,hatch_15,hatch_16,hatch_17,hatch_19,kotschenreuther_19,liu_arxiv_20,chapman_21,hassan_pop_21,guttenfelder_NF_21,parisi_NF_20,parisi_22,walker_23}, where ETG appears to produce experimentally relevant transport levels in many scenarios.  Several recent papers have formulated reduced models and/or simple algebraic expressions for ETG transport in the pedestal~\cite{chapman_21,guttenfelder_NF_21,hatch_22,farcas_22}.  This paper presents simple, yet accurate, algebraic expressions for ETG transport in the pedestal improving on those described in Ref.~\cite{hatch_22}.  These expressions are used to model electron temperature pedestal profiles in representative experimental discharges from DIII-D, JET, and Alcator C-Mod.  

The results of this paper may be distilled into three main conclusions.  First, ETG transport in the pedestal exhibits a surprising insensitivity to geometry, likely due to its predominantly slab-like nature.  The model is based on a database of nonlinear gyrokinetic simulations, described in Ref.~\cite{hatch_22}, of multiple machines (DIII-D, C-Mod, JET) and operating scenarios (e.g., ELMy H-mode, I-mode, carbon walls, metal walls).  Here we show that simulations for spherical tokamaks also fall in line with the model as long as the gradients are sufficiently steep.  Second, the ETG model is characterized by extremely strong dependence on electron temperature (destabilizing) and density (stabilizing) gradients.  This translates into extreme sensitivity of the modeled temperature profiles to separatrix density and temperature, with potentially profound implications for core-edge integration.  Third, ETG transport is likely the main electron thermal transport mechanism in high $\eta = L_n/L_T$ scenarios  like I-mode and JET with an ITER-like wall (ILW) ($L_n$ and $L_T$ are the density and temperature gradient scale lengths, respectively).  In lower $\eta_e$ scenarios, an additional electron thermal transport mechanism is likely necessary to account for the transport.  This is consistent with transport from microtearing modes, which has been identified and investigated in these low-$\eta_e$ scenarios~\cite{hatch_21,hassan_NF_21}.  We present a proof-of-concept MTM model and demonstrate that, in combination with the ETG model, it can reproduce the $T_e$ profile in a DIII-D discharge.  Based on these results, we propose that pedestal predictions of future devices should be tested for compatibility with ETG transport.

The paper is outlined as follows.  In Sec.~\ref{sec:model}, the improved model is defined and compared with a database of nonlinear gyrokinetic simulations. In Sec.~\ref{sec:additional_data}, we show that data from spherical tokamak pedestal simulations follows the same formula and identify limits of applicability of the model.  In Sec.~\ref{sec:comparisons}, the ETG formula is used to model pedestal temperature profiles for four experimental discharges taking into account uncertainties in the inputs.  In Sec.~\ref{sec:MTM}, a proof-of-concept MTM model is introduced and applied to a DIII-D discharge.  Summary and discussion is provided in Sec.~\ref{sec:summary}. 

\section{Improved model for ETG transport}
\label{sec:model}

The refined formula for ETG transport in the steep gradient region of the pedestal is shown in Eq.~\ref{eq:model1}:
\begin{equation}
 \label{eq:model1}
Q_e/Q_{GB} = a_0 \sqrt{m_e/m_i} \omega_{Te}^2 (\eta_e - 1) \eta_e^{b_0} \tau^{c_0} \lambda_D^{2 d_0} 
\end{equation}
with $a_0 = 0.019$, $b_0 = 1.57$, $c_0 = -0.5$, and $d_0 = -0.2$.  In this equation, $Q_e$ is the electron heat flux, $\omega_{Te} = \frac{1}{T_e}\frac{d T_e}{ d \psi}$, $\omega_{ne} = \frac{1}{n_e}\frac{d n_e}{ d \psi}$, $\eta_e = \omega_{Te}/\omega_{ne}$, $\tau = Z_{eff} T_e/T_i$, $\lambda_D$ is the Debye length normalized to $\rho_s = \sqrt{T_e/m_i} \frac{m_i}{e B}$, and the gyroBohm heat flux is $Q_{GB}=n_e T_e c_s \rho_s^2/a^2$, $a$ is the minor radius defined here as $a = \sqrt{2 |\Phi_{edge}|/B_0}$, $\Phi_{edge}$ is the separatrix toroidal flux, and $c_s = \sqrt{T_e/m_i}$.  The radial coordinate, $\psi$, is the normalized poloidal flux.  Note that a more natural gyroBohm normalization, $Q_{GB2} = n_e T_e v_{Te} \rho_e^2/L_{Te}^2$, has sometimes been used in the literature~\cite{hatch_15,chapman_21,guttenfelder_NF_21}.  However, we retain the expression, $Q_{GB}$, defined above since it is the normalization used in the GENE code and it highlights the strong dependence on $\omega_{Te}$ more explicitly.  

Eq.~\ref{eq:model1} will be used in the remainder of the paper.  However, since the dependence on the Debye length is rather weak, we also define a model without it for reference:
\begin{equation}
 \label{eq:model2}
Q_e/Q_{GB} = a_0 \sqrt{m_e/m_i}\omega_{Te}^2 (\eta_e - 1) \eta_e^{b_0} \tau^{c_0}
\end{equation}
with $a_0 = 0.1$, $b_0 = 1.54$, and $c_0 = -0.5$.

The parameters $a_0$, $b_0$, $c_0$, and $d_0$ were numerically optimized to minimize model error in comparison with the database of nonlinear simulation data described in Ref.~\cite{hatch_22}.  We briefly outline, conceptually, the main features of the model.  The most fundamental aspect of the model is the gyroBohm scaling to which the heat flux is normalized.  On the RHS of Eqs.~\ref{eq:model1} and ~\ref{eq:model2}, one factor of $\omega_{Te}$ comes from the electron temperature gradient in Fick's law: $Q \sim n \nabla T \chi$.  The other factor of $\omega_{Te}$ comes from the scaling of the relevant turbulent timescale $v_{Te}/L_{Te}$, where $v_{Te}$ is the electron thermal velocity.  The term $\eta_e-1$ represents a linear (or nonlinear) threshold and will be discussed in more depth below.  The additional $\eta_e^{b_0}$ dependence likely represents some aspect of the nonlinear saturation.  This claim is based on Ref.~\cite{hatch_22}, which describes a quasilinear model for ETG transport that requires additional factors of $\eta_e$ to match nonlinear simulation data, suggesting that the strong $\eta_e$ dependence is a nonlinear effect.  The $\tau$ and $\lambda_D$ dependences correspond to qualitatively well-know trends of the underlying ETG instability.  

\begin{figure}[H]
    \centering
    \includegraphics[scale=0.8]{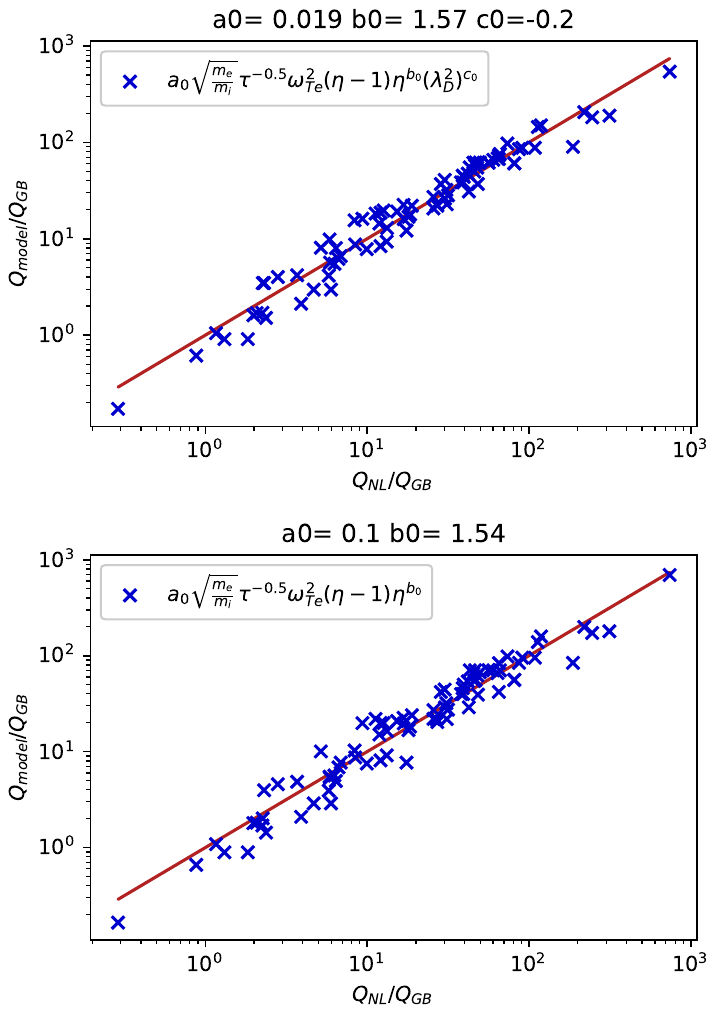}
    \caption{\label{new_models}  The models defined in Eqs.~\ref{eq:model1} (top) and~\ref{eq:model2} (bottom) plotted against nonlinear simulation data. The errors are $15\%$ and $17\%$ respectively. The red line represents perfect agreement between the model and the simulation data.    }
\end{figure}

The results of these models are shown in Fig.~\ref{new_models} plotted against a database of nonlinear gyrokinetic simulations described in Ref.~\cite{hatch_22}.  Model errors are quantified with the expression in Eq.~\ref{error}: 
\begin{equation}
   \varepsilon =  \sqrt{\frac{1}{N} \sum \frac{\left ( Q_{NL} - Q_{model} \right )^2}{ ( Q_{model} + Q_{NL})^2} }, 
\label{error}
\end{equation}
The error for Eq.~\ref{eq:model1} is $\varepsilon = 0.15$ and for Eq.~\ref{eq:model2}, $\varepsilon = 0.17$.  The accuracy of Eq.~\ref{eq:model1} approaches the accuracy of the quasilinear model described in Ref.~\cite{hatch_22} but without the need for linear gyrokinetic simulations.    

Eqs.~\ref{eq:model1} and ~\ref{eq:model2} have been modified in several ways from Ref.~\cite{hatch_22}.  First, all quantities have been formulated in terms of the normalized poloidal flux, $\psi$, which is better behaved in proximity to the separatrix, as opposed to the square root of the normalized toroidal flux $\rho_{tor}$.  In addition, the heat flux, $Q = P/A$ where $P$ is the power, uses the flux surface area $A = 2 \pi R 2 \pi a$, where $R$ is the major radius and $a = \sqrt{2 |\Phi_{edge}|/B_0}$ is a measure of the minor radius, $\Phi_{edge}$ is the toroidal flux at the separatrix, and $B_0$ is the on-axis magnetic field.  This measure of the flux surface area is less sensitive to flux surface shape than $V^\prime$, which was used in Ref.~\cite{hatch_22}.  Both of these changes reduce the impact of geometry and flux surface shaping on the relevant quantities.  To provide a more explicity example, the quantity $V^\prime$ depends sensitively on the flux surface shape, which propagates into the quantity $Q$ calculated by GENE.  We desire a simple expression for the flux that does not depend sensitively on flux surface shape.  One option would be to translate fluxes directly into Watts---i.e. define it in terms of power rather than power per unit area.  We opt instead for a definition of flux surface area that is less dependent on shaping as defined above.   

\begin{figure}[H]
    \centering
    \includegraphics[scale=0.8]{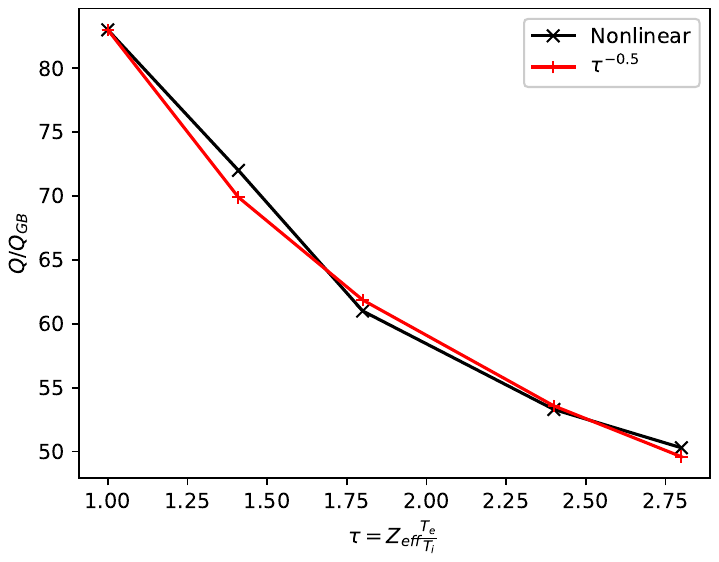}
    \caption{\label{Q_vs_tau} Heat fluxes from a nonlinear gyrokinetic scan of $\tau = Z_{eff} T_e/T_i$.    }
\end{figure}

\begin{figure}[H]
    \centering
    \includegraphics[scale=0.8]{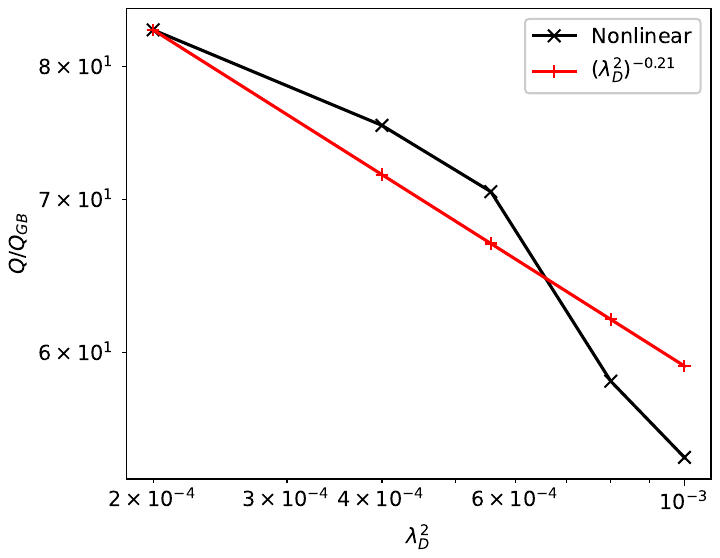}
    \caption{\label{Q_vs_debye2} Heat fluxes from a nonlinear gyrokinetic scan of $\lambda_{D}$.   }
\end{figure}

The dependences on $\tau$ and $\lambda_D$ have also been refined via dedicated scans of these parameters with nonlinear GENE~\cite{jenko_00b,goerler_11} simulations shown in Figs.~\ref{Q_vs_tau} and~\ref{Q_vs_debye2}.  The scalings $\tau^{-0.5}$ and $\lambda_D^{-0.21}$ were determined by optimization with respect to the nonlinear simulation database and are in good or reasonable agreement with the parameter scans. 

Additional modifications from Ref.~\cite{hatch_22} are of note.  The exponent on $\eta_e$ is no longer restricted to integers with the best fit being $\eta_e^{\sim 1.5}$.  We also implement a rough threshold dependence in the term $\eta_e-1$.  The precise parameter dependences of the threshold are not determined, nor is the question of the extent to which a nonlinear threshold agrees with a linear threshold.  However, theory and simulation identify a threshold in the range of $\eta_e = 1$~\cite{jenko_01b}, which we use in the model.  We note that the experimentally-relevant parameter points are typically well beyond this threshold in the pedestal (see Fig. $1$ in Ref.~\cite{hatch_22}), which mitigates sensitivity to the precise threshold value.  An additional difference from Ref.~\cite{hatch_22} is the use of $\omega_{Te}$ to the second instead of first power, which is a better fit given the other modifications described above.  We note that this model still retains extremely strong dependence of the transport on the gradients: $\omega_{Te}$ and $\eta_e$.  

\subsection{Additional Data Points and Limits of Applicability}
\label{sec:additional_data}

\begin{figure}[H]
    \centering
    \includegraphics[scale=0.8]{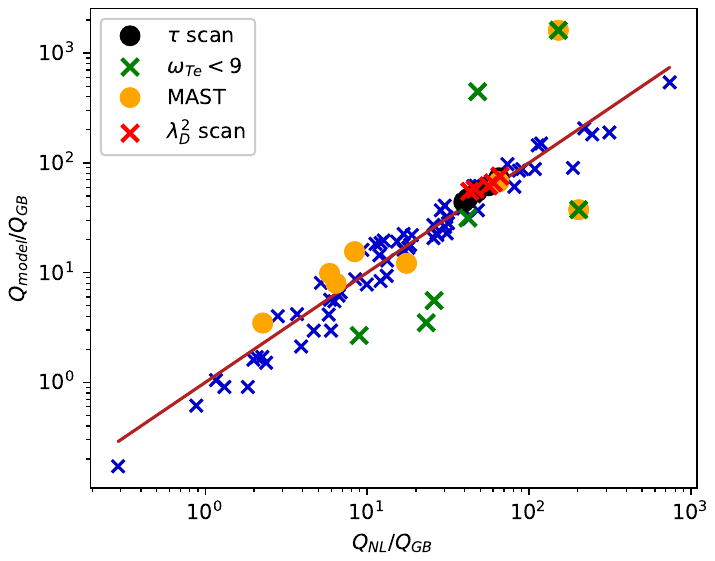}
    \caption{\label{model_with_new_data}  Eq.~\ref{eq:model1} plotted against nonlinear simulation data including some points not shown in Fig.~\ref{new_models}.  Orange symbols represent simulations of the MAST pedestal~\cite{pingyu_23}.  Black and red symbols correspond to the $\tau$ and $\lambda_D$ scans, respectively.  All parameters points with $\omega_{Te} < 9$ are denoted with a green x.  These points come from simulations at the pedestal top where toroidal ETG is more pronounced and Eq.~\ref{eq:model1} is less accurate.  All other points are captured quite accurately by Eq.~\ref{eq:model1}.       }
\end{figure}

Fig.~\ref{model_with_new_data} replicates the data in Fig.~\ref{new_models} with some data points highlighted and some additional data points plotted.  The black and red symbols denote the dedicated $\tau$ and $\lambda_D$ scans, respectively.  These scans would follow horizontal lines in the absence of the additional parameter dependences in Eq.~\ref{eq:model1}.

The additional data is marked with orange circles and green crosses.  The orange symbols represent local nonlinear GENE simulations of the MAST pedestal reported in Ref.~\cite{pingyu_23}.  Notably, the model captures the data quite well with the exceptions being low gradient pedestal top parameter points denoted with green symbols, which will be discussed below.  This agreement is striking considering the rather extreme differences in magnetic geometry, MAST being a spherical tokamak and all other data coming from standard aspect ratio tokamaks.  We note also the diversity in the rest of the dataset, which includes simulations from multiple devices (DIII-D, JET, C-Mod) and multiple scenarios (e.g., ELMy H-mode, I-mode, carbon walls, metal walls).  See Ref.~\cite{hatch_22} for the distribution of meaningful parameters.    

The green symbols denote points with $\omega_{Te} < 9$, which were not shown in Fig.~\ref{new_models}.  The green symbols that coincide with orange circles are simulations at the pedestal top in MAST while the other green symbols denote pedestal top simulations in standard aspect ratio tokamaks.  The cutoff $\omega_{Te}<9$ was chosen rather arbitrarily due to its ability to discriminate between parameter points where the model is or is not accurate.  As gradients decrease in the pedestal top, the toroidal drift frequencies, whose characteristic scale length is the machine size, become increasingly comparable to $\omega_*$ and toroidal drift resonances become more important.  There are additional nuances~\cite{chapman_23} to the discussion of slab and toroidal ETG instabilities, which will be discussed in the next section.  We conclude that the model is most accurate in the steep gradient region where slab ETG is predominant with degrading accuracy toward the pedestal top.   

We briefly note additional interesting effects that have been reported in the literature and which may not be fully (or even partially) captured by this model: nonlocal effects on ETG transport due to profile curvature at the pedestal top~\cite{pingyu_23}, toroidal ETG modes at low $k_y \rho_e$~\cite{parisi_NF_20,parisi_22}, and cross-scale coupling~\cite{pueschel_20,belli_23}.  


\section{Comparisons with other models}
\label{sec:comparisons}

In this section we compare Eq.~\ref{eq:model1} with other formulas in the literature.  Eq.~\ref{chapman_model} was derived from a fit to an $\eta_e$ scan around a parameter point from a JET-ILW discharge as reported in Ref.~\cite{chapman_21}.  It was then applied to modeling JET-ILW temperature profiles in Ref.~\cite{field_23}~\footnote{Although the formula in Ref.~\cite{field_23} used a different expression for the thermal velocity than that used to derive Eq. 4 in Ref.~\cite{chapman_21}, which uses the GENE normalization}.  
\begin{equation}
\label{chapman_model} 
Q/Q_{GB} =  a_0 \sqrt{\frac{m_e}{m_i}} \omega_{Te}^2 \left ( \omega_{Te} - 1.24 \omega_{ne} \right )^{1.3}
\end{equation}
Taking this functional form, the best fit to the gyrokinetic database is $a_0 = 0.012$, which results in an error $\varepsilon = 0.48$.  

The formula in Eq.~\ref{guttenfelder_model} was derived from several $\eta_e$ scans based on two DIII-D discharges as described in Ref.~\cite{guttenfelder_NF_21}.  
\begin{equation}
\label{guttenfelder_model} 
Q/Q_{GB} =  a_0 \sqrt{\frac{m_e}{m_i}} \omega_{Te}^2 \left ( \omega_{Te}/\omega_{ne} - 1.4 \right )
\end{equation}
With this functional form, the best fit to the gyrokinetic database is  $a_0 = 0.6$, which results in $\varepsilon = 0.31$.

\begin{figure}[H]
    \centering
    \includegraphics[scale=0.8]{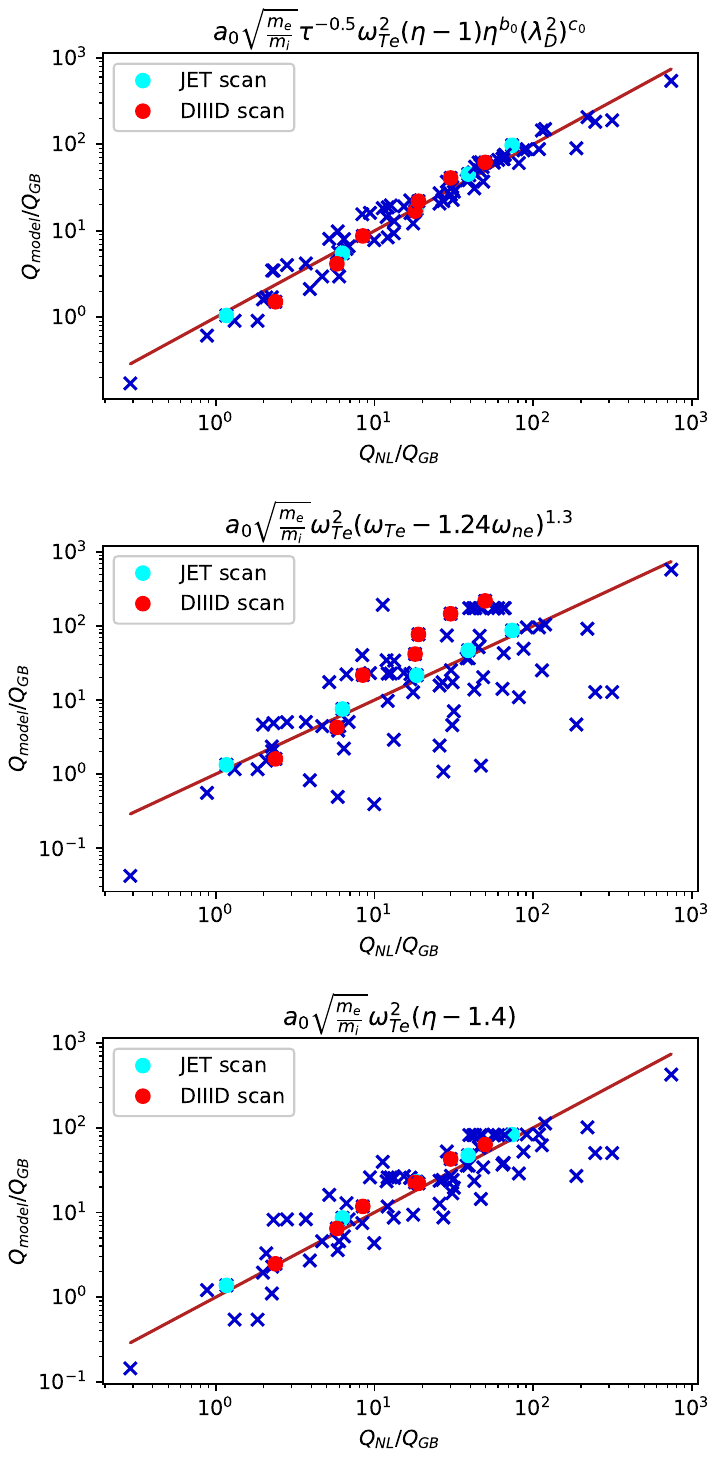}
    \caption{\label{model_comparison}  Eq.~\ref{eq:model1} (top), the model in Ref.~\cite{chapman_21} (middle), and the model in Ref.~\cite{guttenfelder_NF_21} (bottom) plotted against the database of nonlinear simulations.  Cyan points are from Ref.~\cite{chapman_21} and red points are from Ref.~\cite{guttenfelder_NF_21}. }
\end{figure}

Comparisons of the models are shown in Fig.~\ref{model_comparison}.  In this figure, the bulk of the database is shown in blue symbols, the data used to formulate Eq.~\ref{chapman_model} (the left plot in Fig. 12 of Ref.~\cite{chapman_21}) is shown in cyan, and the data used for Eq.~\ref{guttenfelder_model} is shown in red symbols (from the top plot of Fig. $8$ in Ref.~\cite{guttenfelder_NF_21}).   Fig.~\ref{model_comparison} shows the results of Eq.~\ref{eq:model1} in the top panel, the results of Eq.~\ref{chapman_model} in the middle panel, and the results of Eq.~\ref{guttenfelder_model} in the bottom panel.  Eqs.~\ref{chapman_model} and ~\ref{guttenfelder_model} reproduce their respective $\eta_e$ scans very well but overall exhibit higher error throughout the database.  While Eq.~\ref{eq:model1} does not reproduce each $\eta_e$ scan as accurately, they all collapse into a narrow region and there is lower error throughout the database.  The distinguishing feature of Eq.~\ref{eq:model1}, in comparison with the other models, is the higher powers of $\eta_e$.  

Thus far we have framed the discussion of pedestal ETG instabilities as a transition from toroidal ETG at the pedestal top to slab ETG in the steep gradient region.  We believe this picture is basically accurate.  However, Ref.~\cite{chapman_23} introduces interesting nuances to this story that may be connected to the differences in $\eta$ scalings defined in Eqs.~\ref{eq:model1},~\ref{chapman_model},~\ref{guttenfelder_model}.  Ref.~\cite{chapman_23} compares two JET-ILW scenarios, clearly demonstrating that the moderate $\eta_e$ scenario (low gas, high $\omega_n$) is completely slab dominated while a high $\eta_e$ scenario (high gas, weaker $\omega_n$) has an additional component of small scale toroidal ETG (not the low-$k_y$ toroidal modes of Ref.~\cite{parisi_NF_20}).  This motivates a speculative hypothesis that the stronger $\eta_e$ scaling in Eq.~\ref{eq:model1} parameterizes the $\eta_e$ dependence of this additional component of toroidal ETG transport.  Further investigation of this hypothesis must be left for future work.

We proceed in the next section to test profile predictions with Eq.~\ref{eq:model1}.  

\section{Experimental Comparisons}
\label{sec:experimental}

In this section, we apply Eq.~\ref{eq:model1} to modeling electron temperature profiles from various experimental scenarios.  Similar exercises were carried out in Ref.~\cite{field_23}, which showed that JET-ILW profiles can be reproduced using the ETG formulas in Ref.~\cite{chapman_21}.~\footnote{We note that this paper used an expression for the thermal velocity that results in larger transport in comparison with the gyrokinetic data in Ref.~\cite{chapman_21}.}  Ref.~\cite{guttenfelder_NF_21} presented an ETG model (as described in the previous subsection) and applied it to two DIIID discharges, concluding that additional electron thermal transport is likely required to reproduce the profiles.  

\begin{figure}[H]
    \centering
    \includegraphics[scale=0.8]{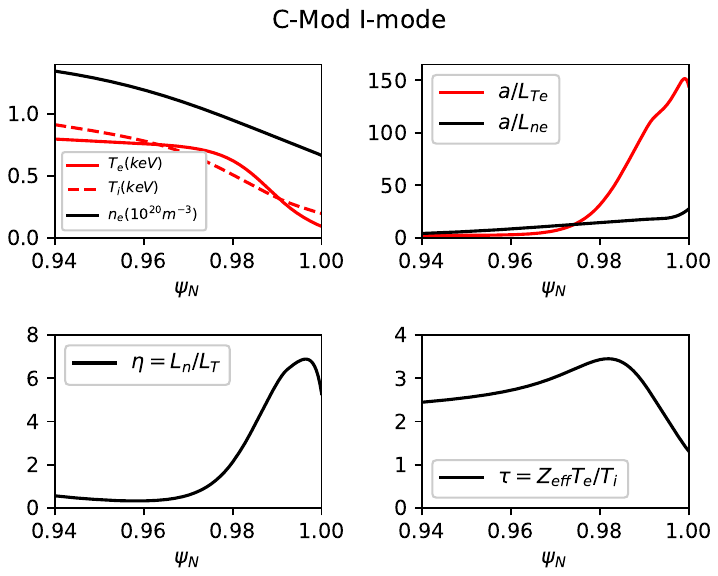}
    \caption{\label{Imode_profiles}  Temperature and density profiles (top left), $a/L_{Te}$ and $a/L_{ne}$ (top right), $\eta_e$ (bottom left), and $\tau$ (bottom right) for the I-mode pedestal from C-Mod discharge 1120907032.    }
\end{figure}

Here we investigate four experimental scenarios selected to probe various aspects of the model and pedestal transport.  Perhaps the most distinguishing feature of Eq.~\ref{eq:model1} is the strong dependence on gradients: $\omega_{Te}^2$ and $\eta_e^{\sim 2.5}$.  This motivates an examination of the I-mode, which is an intrinsically high-$\eta_e$ pedestal scenario.  Here, we investigate the I-mode pedestal from C-Mod discharge 1120907032, for which a gyrokinetic study was reported in Ref.~\cite{liu_arxiv_20}.  If ETG constrains the electron temperature pedestal, one would expect this to be clearly manifest in such a high-$\eta_e$ scenario.  Moreover, if the strong $\eta_n$ dependence of Eq.~\ref{eq:model1} is spurious, this scenario offers a strong test as well.  Profiles for this discharge are shown in Fig.~\ref{Imode_profiles}.  Note the extreme values of $\eta_e$ approaching $\eta_e = 7$.   

In addition, we examine three ELMy H-mode scenarios.  JET discharges 92432 and 78697 were studied in Ref.~\cite{hatch_19} as a comparison between ITER-like wall (ILW) pedestals and earlier carbon (C) wall pedestals.  The JET-ILW pedestals generally exhibit higher $\eta_e$ due to the effect of gas puffing on the density pedestal, which is required to ameliorate sputtering from the tungsten divertor surface~\cite{hatch_17,hatch_19,frassinetti_17,frassinetti_20,frassinetti_21}.  The profiles for the JET-ILW discharge are shown in Fig.~\ref{JET92432_profiles} and the profiles for the JET-C discharge are shown in Fig.~\ref{JET78697_profiles}.    

\begin{figure}[H]
    \centering
    \includegraphics[scale=0.8]{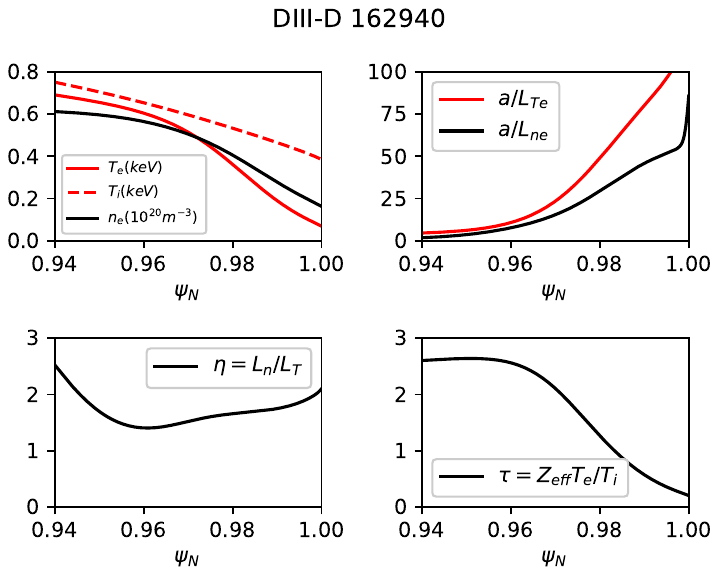}
    \caption{\label{DIIID_profiles}  Temperature and density profiles (top left), $a/L_{Te}$ and $a/L_{ne}$ (top right), $\eta_e$ (bottom left), and $\tau$ (bottom right) for the DIII-D discharge 162940.  }
\end{figure}

\begin{figure}[H]
    \centering
    \includegraphics[scale=0.8]{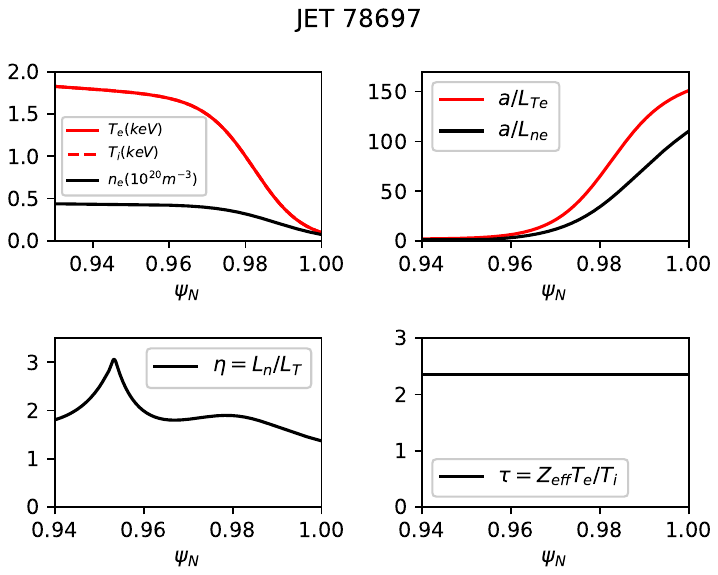}
    \caption{\label{JET78697_profiles}   Temperature and density profiles (top left), $a/L_{Te}$ and $a/L_{ne}$ (top right), $\eta_e$ (bottom left), and $\tau$ (bottom right) for the JET discharge 78697. }
\end{figure}

\begin{figure}[H]
    \centering
    \includegraphics[scale=0.8]{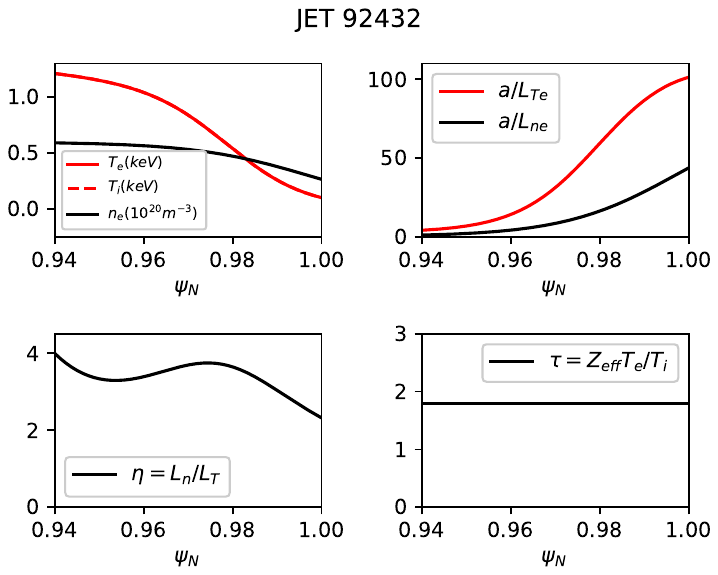}
    \caption{\label{JET92432_profiles}  Temperature and density profiles (top left), $a/L_{Te}$ and $a/L_{ne}$ (top right), $\eta_e$ (bottom left), and $\tau$ (bottom right) for the JET discharge 92432. }
\end{figure}

We also investigate DIII-D discharge 162940, which has been the object of several pedestal gyrokinetic studies~\cite{guttenfelder_NF_21,hassan_pop_21,hassan_NF_21}.  Profiles are shown in Fig.~\ref{DIIID_profiles}.  One motivation for studying DIII-D 162940 and JET 78697 is that microtearing modes (MTM) have been identified in these discharges and studied extensively using gyrokinetics~\cite{hatch_19,hatch_21,hassan_NF_21}.  In both of these discharges, gyrokinetic simulations were able to reproduce distinctive signatures in the frequency spectra of the magnetic fluctuations.  The transport signature of MTMs is identical to that of ETG transport: they both produce almost exclusively electron thermal transport.  If MTMs are indeed a major transport mechanism in these discharges, one would expect the ETG model proposed here to over-predict the temperature pedestal, which turns out to be the case as described below.  

\begin{figure}[H]
    \centering
    \includegraphics[scale=0.8]{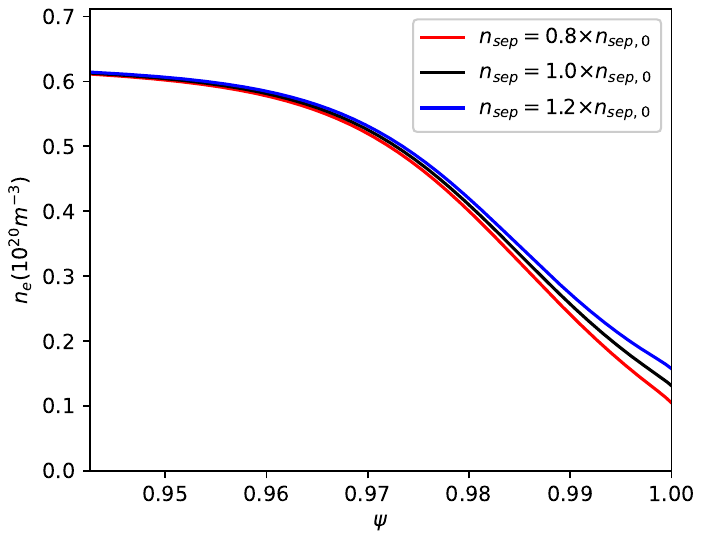}
    \caption{\label{nsep_plot}  Density profiles modified to keep $n_e$ fixed at the pedestal top while varying separatrix density by $\pm 20\%$.  }
\end{figure}

\begin{figure}[H]
    \centering
    \includegraphics[scale=0.8]{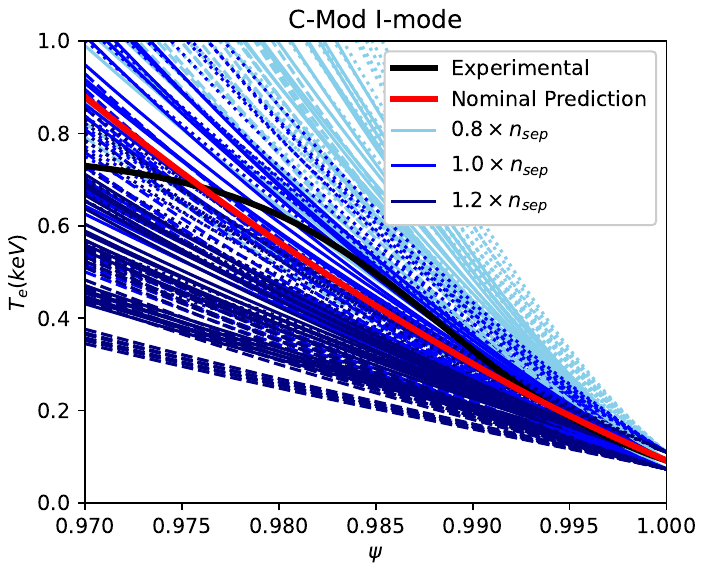}
    \caption{\label{Imode}  The experimental electron temperature profile for I-mode discharge 1120907032 (black) along with the modeled profile with the nominal experimental inputs (red), and modeled profiles resulting from variations in power, $T_{e,sep}$, $n_{sep}$, $Z_{eff}$, and $T_i$.    }
\end{figure}

\begin{figure}[H]
    \centering
    \includegraphics[scale=0.8]{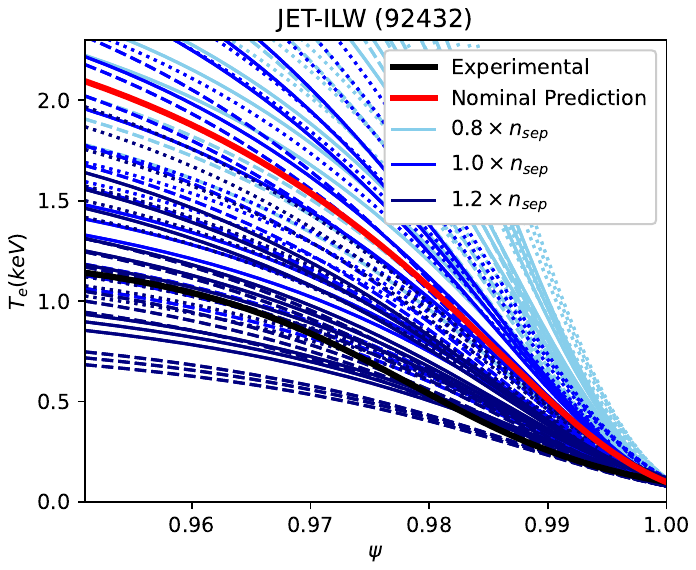}
    \caption{\label{JET92432}  The experimental electron temperature profile for JET discharge 92432 (black) along with the modeled profile with the nominal experimental inputs (red), and modeled profiles resulting from variations in power, $T_{e,sep}$, $n_{sep}$, $Z_{eff}$, and $T_i$. }
\end{figure}

\begin{figure}[H]
    \centering
    \includegraphics[scale=0.8]{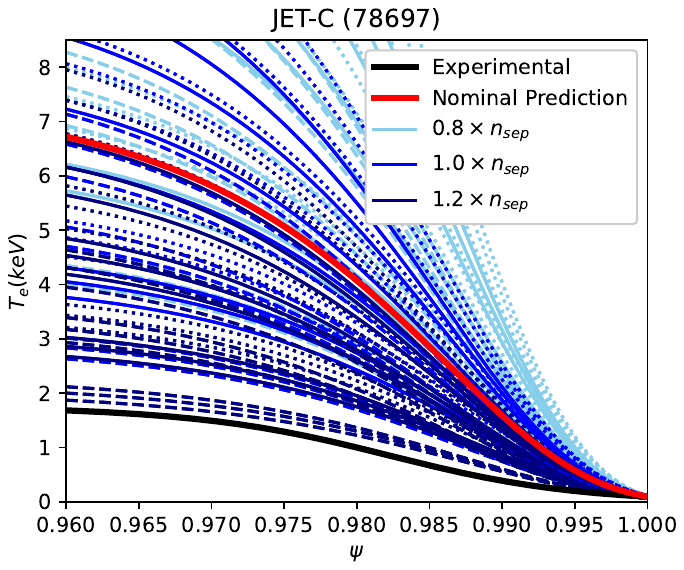}
    \caption{\label{JET78697} The experimental electron temperature profile for JET discharge 78697 (black) along with the modeled profile with the nominal experimental inputs (red), and modeled profiles resulting from variations in power, $T_{e,sep}$, $n_{sep}$, $Z_{eff}$, and $T_i$.  }
\end{figure}

\begin{figure}[H]
    \centering
    \includegraphics[scale=0.8]{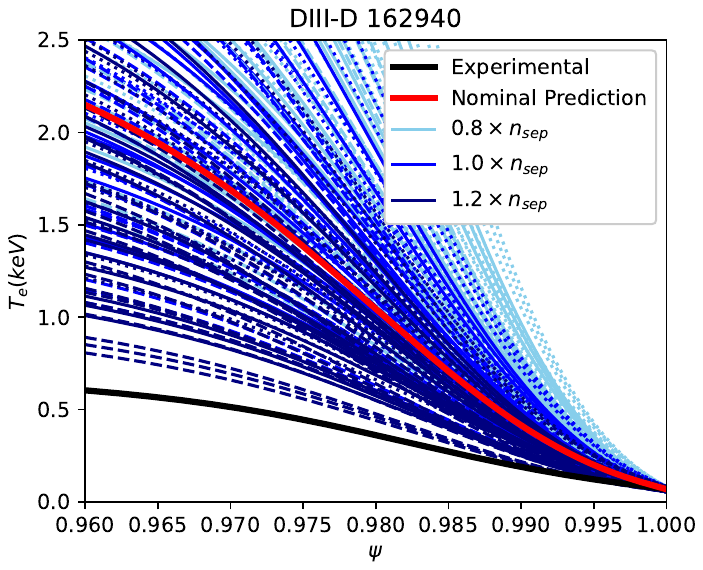}
    \caption{\label{DIIID162940}  The experimental electron temperature profile for DIII-D discharge 162940 (black) along with the modeled profile with the nominal experimental inputs (red), and modeled profiles resulting from variations in power, $T_{e,sep}$, $n_{sep}$, $Z_{eff}$, and $T_i$. }
\end{figure}

The electron temperature profiles are modeled holding all other profiles fixed.  The profiles were solved numerically by starting from the separatrix and integrating inward using Eq.~\ref{eq:model1} to determine the value of $\omega_{Te}$ that satisfies power balance at each radial position.  The inputs are (1) separatrix electron temperature, (2) the ion temperature profile, (3) the electron density profile, (4) the profile of $\tau = Z_{eff} T_e/T_i$, (5) the electron heat flux through the pedestal (assumed to be radially constant), and (6) the magnetic field, and major/minor radii.  In order to probe sensitivities, we vary combinatorially the flux by $\sim \pm 50\%$ (since there is substantial uncertainty in the distribution of the heat flux between electrons and ions), the electron separatrix density by $\pm 20\%$ (an example is shown in Fig.~\ref{nsep_plot}), the separatrix electron temperature by $\pm20\%$, and $Z_{eff}$ by $\pm20\%$.  We also iterate between two bounds for the ion temperature: the experimentally-estimated value (if available) and setting $T_i=T_e$.  The total power (ions and electrons) entering the pedestal is $4$ MW for the I-mode discharge~\cite{liu_arxiv_20}, $3$ MW for the DIII-D discharge~\cite{guttenfelder_NF_21}, $11.6$ MW for the JET-ILW discharge~\cite{field_18}, and $5.7$ MW for the JET-C discharge~\cite{field_18}.  The resulting profiles are extremely sensitive to the electron separatrix density, which is to be expected from the strong dependence on $\eta_e$ in Eq.~\ref{eq:model1}.  Such sensitivity to separatrix boundary conditions has also been noted in Refs.~\cite{hatch_17,kotschenreuther_17,frassinetti_21,field_23}.  Results are shown in Figs.~\ref{Imode},~\ref{JET92432},~\ref{JET78697},~\ref{DIIID162940}.  In these figures, the experimental profiles are shown in black, the profiles predicted with the nominal experimental inputs are shown in red, and the variations are shown in blue.  The color coding highlights the impact of the strongest parameter dependence: lighter to darker blue as the separatrix density progresses from $-20\%$ to $+20\%$.

Several observations can be made.  First, the model reproduces the I-mode temperature profile quite accurately: the nominal prediction matches the experimental profiles quite well and the experimental profile lies firmly in the spread of predicted profiles from the sensitivity tests.  We conclude from this that the model is likely quite accurate in the high-$\eta_e$ regime.  Moreover, we posit that the electron temperature profile for this I-mode discharge is likely set almost entirely by ETG transport.  Despite the good agreement in the steep gradient region of the pedestal, the model does not predict a flattening of the $T_e$ profile at the pedestal top---i.e., the model does not produce the familiar $tanh$ profile shape.  This is different from the conventional H-mode scenarios discussed below.  The reason for this is that the I-mode density profile does not flatten in this region in contrast with the other H-mode scenarios.  This discrepancy between modeled and observed $T_e$ at the pedestal top could be resolved by either (1) some additional electron thermal transport mechanism in this region possibly at ion scales, or (2) perhaps an improved ETG model that captures the transition to toroidal ETG at the pedestal top.

The other discharges studied here are ELMy H-modes, for which kinetic ballooning modes (KBM) are thought to constrain the pressure profile~\cite{snyder_09}.  We add a rough estimate of the contribution to the electron heat flux from KBM.  Basic theory and gyrokinetic simulations~\cite{kotschenreuther_19} predict the ratio $D_e/\chi_e~\sim 2/3$ for KBMs.  Interpretive edge modeling generally predicts $D_e/\chi_e \sim 0.1-0.3$ in the pedestal (references).  Hence we proceed with an upper-bound estimate that KBM can account for half of the electron thermal transport in the pedestal.  Note that we do not apply this additional transport mechanism for the I-mode modeling described above since I-mode pressure profiles lie below the KBM limit.  As of now we neglect any possible contribution from MTM, but this will be addressed in the next section.  

The modeled $T_e$ profile for the JET-ILW scenario is shown in Fig.~\ref{JET92432}.  The model generally over predicts the temperature profile, although some of the parameter variations produce temperature profiles in the right range.  Based on this model, ETG transport plausibly determines the $T_e$ profile.  However, the results also suggest that additional electron heat transport mechanisms may be active in the pedestal.  Ref.~\cite{hatch_19} predicted a non-negligible contribution to the electron thermal transport from ion scales---almost $2MW$ for some simulations.   

For the DIII-D pedestal and the JET-C pedestal the ETG model over-predicts the electron temperature even when taking into account the parameter variations.  This is consistent with an additional contribution to the transport from MTM, which has been studied in detail via gyrokinetic simulations for both of these discharges and connected directly with experimental fluctuation data~\cite{hatch_21,hassan_NF_21}.  MTM transport will be probed in more depth in the next section.  Despite the failure of the ETG model to predict (single-handedly) the electron temperature profiles, nonlinear gyrokinetic simulations of ETG transport based directly on the nominal experimental parameters have been shown in previous studies to produce experimentally-relevant transport levels for all four of these discharges~\cite{hatch_19,liu_arxiv_20,hassan_pop_21,guttenfelder_NF_21}.  Consequently, we posit that ETG likely plays a role in the electron thermal transport, particularly in the upper pedestal, even in scenarios where it cannot reproduce the entire temperature profile.  


\section{Proof of Concept MTM Models}
\label{sec:MTM}

As described in the previous section, an additional electron thermal transport mechanism is likely necessary to model pedestal temperature profiles in lower-$\eta_e$ pedestal scenarios.  Here we explore a proof-of-concept model for pedestal MTM transport.  We emphasize that this model should not be considered in the same class as the ETG model described above, which is based on an extensive database of first-principles nonlinear gyrokinetic simulations.  Rather, this is simply a demonstration (1) of how such models may be developed in future work, and (2) that plausible models for MTM can reproduce $T_e$ profiles in combination with the ETG model.

We begin with the following ansatz: (1) the underlying scaling of MTM transport is gyroBohm, (2) MTM is unstable above a critical gradient in $\omega_{Te}$, and (3) it is unstable at the peak of the $\omega_*$ `well'~\cite{hatch_21,hassan_NF_21,larakers_21,curie_22}, which is characteristic of the pedestal, and has some radial envelope.  This ansatz is quantified in Eq.~\ref{eq:generalized_mixing}:
\begin{equation}
 \label{eq:generalized_mixing}
Q = a_0 Q_{GB} \omega_{Te}^{b_0} \left ( \frac{\omega_{Te}}{\omega_{Te,crit}}-1 \right ) e^{-(\psi_N-\psi_*)^2/d\psi^2}
\end{equation}

\begin{figure}[htb!]
    \centering
    \includegraphics[scale=0.8]{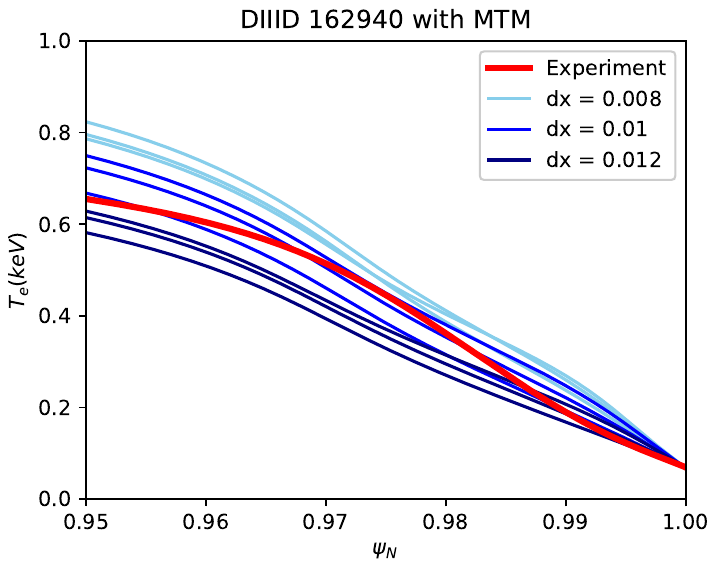}
    \caption{\label{DIIID_w_MTM}  Experimental $T_e$ profile from DIII-D discharge 162940 (red) along with modeled profiles using the ETG model defined in Eq.~\ref{eq:model1} in combination with the model for MTM transport defined in Eq.~\ref{eq:generalized_mixing} with variations in model parameters.   }
\end{figure}

\begin{figure}[htb!]
    \centering
    \includegraphics[scale=0.8]{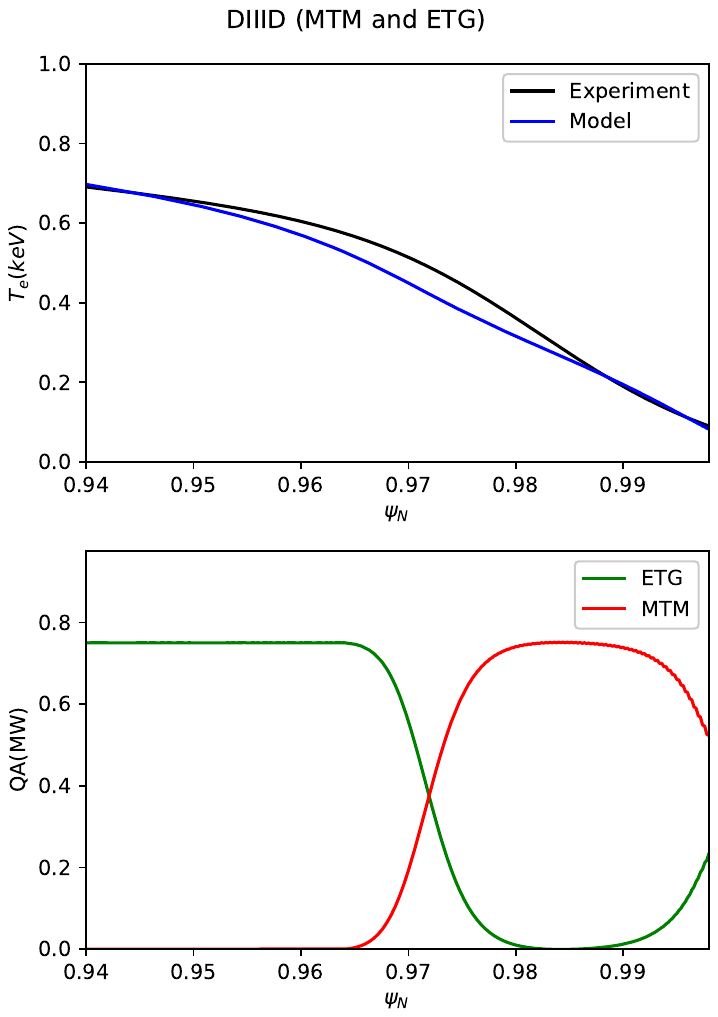}
    \caption{\label{ETG_and_MTM} Top: Experimental $T_e$ profile from DIII-D discharge 162940 along with the modeled $T_e$ profile combining both ETG transport using Eq.~\ref{eq:model1} and MTM transport using Eq.~\ref{eq:generalized_mixing} with model parameters $\omega_{Te,crit} = 22.1$, $a_0 = 3.7 \times 10^{-4}$, $b_0 = 3$.  Bottom: The relative transport from ETG (green) and MTM (red).  ETG produces the pedestal top transport while MTM produces the transport in the steep gradient region.  }
\end{figure}

We apply this to the DIII-D scenario described above, for which the MTM instabilities and transport have been investigated in depth as described in Ref.~\cite{hassan_NF_21}.  The critical gradient $\omega_{Te,crit} = 22.1$ was determined by linear GENE simulations at the peak of the $\omega_*$ profile.  Subsequently, the exponent $b_0=2,3,4$ and the radial width $d\psi=0.008,0.01,0.012$ were scanned.  For each point in the scan, the coefficient $a_0$ was set to satisfy $QA = 4 MW$ for the nominal profile at the peak of the $\omega_*$ profile ($\psi = 0.9818$), which agrees with the nonlinear simulations described in Ref.~\cite{hassan_NF_21}.  This model for MTM transport was combined with the ETG model to predict $T_e$ profiles.  The predicted profiles for this range of parameters are in reasonable agreement with the experimental profile as shown in Fig.~\ref{DIIID_w_MTM}.  As an illustrative example, we further highlight the model parameters $a_0 = 3.7\times10^{-4}$, $b_0= 3$, $d\psi = 0.01$
with $\omega_{Te,crit} = 22.1$, for which the modeled profile is shown in Fig.~\ref{ETG_and_MTM} (top panel) along with the relative contribution of MTM and ETG in the bottom panel.

We emphasize that this MTM model is likely not applicable across a broad parameter space.  MTM stability is much more complex than ETG and depends on a richer set of underlying parameters (notably, collisionality, electron temperature gradient, and $\beta$).  A promising next step would be to formulate a model that directly uses linear gyrokinetic simulations of MTMs within a more rigorous statistical framework informed by both simulation and experimental data.  


\section{Summary and Discussion}
\label{sec:summary}

This paper describes a simple formula for ETG transport in the pedestal, improving on those described in Ref.~\cite{hatch_22} by refining fits for key parameter dependences and reducing the effects of geometric factors.  The model achieves low error in comparison with an extensive database of nonlinear gyrokinetic simulations: $\varepsilon = 15\%$.  The model is compared with other formulas in the literature and found to be more broadly applicable---i.e., lower error---across the database.    

In addition to the standard aspect ratio simulations that represent the bulk of the data, simulations of the spherical tokamak, MAST, also agree with the formula.  It appears that pedestal ETG transport is insensitive to geomtry and the formula is accurate throughout a surprisingly large range of parameter space provided the gradients are sufficiently steep, resulting in predominantly slab-like instabilities.  

The formula for ETG transport is used to model $T_e$ profiles from four experimental discharges.  In all cases, sensitivity to separatrix density results in a broad range of possible $T_e$ profiles.  In the intrinsically high-$\eta_e$ I-mode scenario, the model reproduces the experimental temperature profile quite accurately, suggesting that I-mode is likely constrained by ETG transport.  For a moderately high-$\eta_e$ JET-ILW discharge, the nominal experimental input parameters produce a $T_e$ profile that is somewhat higher than the experimental profile, although the experimental profile does fall well within the range sampled with sensitivity tests of the inputs.  For two low-$\eta_e$ discharges (one from DIII-D and one from JET with a carbon wall), the modeled $T_e$ profiles are well above the experimental profile even taking into account uncertainties in input parameters.  This is consistent with MTM as an additional transport mechanism, which has been identified in previous studies of these discharges.  We explore a simple model for MTM transport and show that for plausible model parameters the combination of MTM and ETG can recover the experimental profile with MTM predominant in the steep gradient region and MTM taking over toward the pedestal top.  Based on these four discharges it appears that ETG is increasingly active as $\eta_e$ increases but MTM is necessary at lower $\eta_e$.  Additional electron thermal transport mechanisms like KBM and/or some transport mechanism near the separatrix like blobby transport or stochastic transport from error fields~\cite{ashourvan_22}.

Pedestal ETG transport is characterized by extremely strong dependence on electron temperature (destabilizing) and density (stabilizing) gradients, which results in extreme sensitivity to separatrix temperature and density.  This is may be related to the observed degradation confinement with increasing separatrix density and decreasing temperature, which has been reported in several recent studies~\cite{frassinetti_17,verdoolaege_21,frassinetti_21,lomanowski_22,bourdelle_23}.  We propose that pedestal predictions for future reactor scenarios should be tested for compatibility with ETG transport.      

{\em Acknowledgements.--} 

We would like to acknowledge useful discussions with Anthony Field.  This research used resources of the National Energy Research Scientific Computing Center, a DOE Office of Science User Facility.  This work was supported by U.S. DOE Contract Nos. DE-FG02-04ER54742, DE-SC0018429, and DE-SC0022164.  
This work has been carried out within the framework of the EUROfusion Consortium, funded by the European Union via the Eu-
ratom Research and Training Programme (Grant No. 101052200 — EUROfusion) and from the RCUK (Grant
No. EP/T012250/1). Views and opinions expressed are however those of the author(s) only and do not necessarily
reflect those of the European Union or the European Commission. Neither the European Union nor the European
Commission can be held responsible for them. 

\section{References}

\bibliography{my_refs}{}

\begin{thebibliography}{10}

\bibitem{hatch_22}
D~R Hatch, C~Michoski, and D~Kuang.
\newblock Reduced models for {ETG} transport in the tokamak pedestal.
\newblock {\em Physics of Plasmas}, page~12, 2022.

\bibitem{kotschenreuther_19}
M.~Kotschenreuther, X.~Liu, D.R. Hatch, S.~Mahajan, L.~Zheng, A.~Diallo,
  R.~Groebner, {the DIII-D TEAM}, J.C. Hillesheim, C.F. Maggi, C.~Giroud,
  F.~Koechl, V.~Parail, S.~Saarelma, E.~Solano, A.~Chankin, and {JET
  Contributors}.
\newblock Gyrokinetic analysis and simulation of pedestals to identify the
  culprits for energy losses using ‘fingerprints’.
\newblock {\em Nucl. Fusion}, 59(9):096001, September 2019.

\bibitem{TPT}
D.~R. Hatch, M.~T. Kotschenreuther, S.~M. Mahajan, M.~Halfmoon, E.~Hassan,
  G.~Merlo, C.~Michoski, J.~Canik, A.~Sontag, I.~Joseph, M.~Umansky,
  W.~Guttenfelder, A.~Diallo, R.~Groebner, A.~O. Nelson, F.~Laggner, J.~Hughes,
  and S.~Mordijck.
\newblock Final report for the fy19 fes theory performance target.
\newblock Technical report, 2019.

\bibitem{jenko_00b}
F.~Jenko, W.~Dorland, M.~Kotschenreuther, and B.N. Rogers.
\newblock Electron temperature gradient driven turbulence.
\newblock {\em Phys. Plasmas}, 7:1904, 2000.

\bibitem{dorland_prl_00}
W.~Dorland, F.~Jenko, M.~Kotschenreuther, and B.~N. Rogers.
\newblock Electron temperature gradient turbulence.
\newblock {\em Phys. Rev. Lett.}, 85:5579--5582, Dec 2000.

\bibitem{told_08}
D.~Told, F.~Jenko, P.~Xanthopoulos, L.~D. Horton, and E.~Wolfrum.
\newblock Gyrokinetic microinstabilities in asdex upgrade edge plasmas.
\newblock {\em Physics of Plasmas}, 15(10):102306, 2008.

\bibitem{jenko_09}
F.~Jenko, D.~Told, P.~Xanthopoulos, F.~Merz, and L.~D. Horton.
\newblock Gyrokinetic turbulence under near-separatrix or nonaxisymmetric
  conditions.
\newblock {\em Physics of Plasmas}, 16(5):055901, 2009.

\bibitem{hatch_15}
D.R. Hatch, D.~Told, F.~Jenko, H.~Doerk, M.G. Dunne, E.~Wolfrum, E.~Viezzer,
  {The ASDEX Upgrade Team}, and M.J. Pueschel.
\newblock Gyrokinetic study of {ASDEX} {Upgrade} inter-{ELM} pedestal profile
  evolution.
\newblock {\em Nuclear Fusion}, 55(6):063028, June 2015.

\bibitem{hatch_16}
D.~R. Hatch, M.~Kotschenreuther, S.~Mahajan, P.~Valanju, F.~Jenko, D.~Told,
  T.~G\"orler, and S.~Saarelma.
\newblock Microtearing turbulence limiting the {JET}-{ILW} pedestal.
\newblock {\em Nuclear Fusion}, 56(10):104003, 2016.

\bibitem{hatch_17}
{D. R. Hatch}, M.~Kotschenreuther, S.~Mahajan, P.~Valanju, and X.~Liu.
\newblock A gyrokinetic perspective on the {JET}-{ILW} pedestal.
\newblock {\em Nuclear Fusion}, 57(3):036020, 2017.

\bibitem{hatch_19}
D.R. Hatch, M.~Kotschenreuther, S.M. Mahajan, G.~Merlo, A.R. Field, C.~Giroud,
  J.C. Hillesheim, C.F. Maggi, C.~Perez~von Thun, C.M. Roach, S.~Saarelma, and
  {JET Contributors}.
\newblock Direct gyrokinetic comparison of pedestal transport in {JET} with
  carbon and {ITER}-like walls.
\newblock {\em Nucl. Fusion}, 59(8):086056, August 2019.

\bibitem{liu_arxiv_20}
Xing Liu, Mike Kotschenreuther, David~R. Hatch, Swadesh~M. Mahajan, Jerry~W.
  Hughes, and Amanda~E. Hubbard.
\newblock Gyrokinetics investigations of an i-mode pedestal on alcator c-mod,
  2020.

\bibitem{chapman_21}
B.~Chapman-Oplopoiou, D.~R. Hatch, A.~R. Field, L.~Frassinetti, J.~C.
  Hillesheim, L.~Horvath, C.~F. Maggi, J.~F. Parisi, C.~M. Roach, S.~Saarelma,
  J.~Walker, and JET contributors.
\newblock The role of etg modes in jet-ilw pedestals with varying levels of
  power and fuelling.
\newblock {\em submitted to Nuclear Fusion}.

\bibitem{hassan_pop_21}
Ehab Hassan, D.~R. Hatch, W.~Guttenfelder, Y.~Chen, and S.~Parker.
\newblock Gyrokinetic benchmark of the electron temperature-gradient
  instability in the pedestal region.
\newblock {\em Physics of Plasmas}, 28(6):062505, June 2021.

\bibitem{guttenfelder_NF_21}
W.~Guttenfelder, R.J. Groebner, J.M. Canik, B.A. Grierson, E.A. Belli, and
  J.~Candy.
\newblock Testing predictions of electron scale turbulent pedestal transport in
  two {DIII}-{D} {ELMy} {H}-modes.
\newblock {\em Nuclear Fusion}, 61(5):056005, May 2021.

\bibitem{parisi_NF_20}
Jason~F. Parisi, Felix~I. Parra, Colin~M. Roach, Carine Giroud, William
  Dorland, David~R. Hatch, Michael Barnes, Jon~C. Hillesheim, Nobuyuki Aiba,
  Justin Ball, Plamen~G. Ivanov, and Jet contributors.
\newblock Toroidal and slab {ETG} instability dominance in the linear spectrum
  of {JET}-{ILW} pedestals.
\newblock {\em Nuclear Fusion}, 60(12):126045, December 2020.

\bibitem{parisi_22}
Jason Parisi and {\it et al.}
\newblock {\em Physics of Plasmas}.

\bibitem{walker_23}
Justin Walker and David~R. Hatch.
\newblock {ETG} turbulence in a tokamak pedestal.
\newblock {\em Physics of Plasmas}, 30(8):082307, August 2023.

\bibitem{farcas_22}
Ionu{\c{t}}-Gabriel Farca{\c{s}}, Gabriele Merlo, and Frank Jenko.
\newblock A general framework for quantifying uncertainty at scale.
\newblock {\em Communications Engineering}, 1(1):43, 2022.

\bibitem{hatch_21}
D.R. Hatch, M.~Kotschenreuther, S.M. Mahajan, M.J. Pueschel, C.~Michoski,
  G.~Merlo, E.~Hassan, A.R. Field, L.~Frassinetti, C.~Giroud, J.C. Hillesheim,
  C.F. Maggi, C.~Perez von Thun, C.M. Roach, S.~Saarelma, D.~Jarema, F.~Jenko,
  and JET Contributors.
\newblock Microtearing modes as the source of magnetic fluctuations in the
  {JET} pedestal.
\newblock {\em Nuclear Fusion}, 61(3):036015, feb 2021.

\bibitem{hassan_NF_21}
Ehab Hassan, David~R Hatch, Michael Halfmoon, Max Curie, Michael
  Kotschenreuther, Swadesh~M Mahajan, Gabriele Merlo, Richard~J Groebner,
  Andrew~Oakleigh Nelson, and Ahmed Diallo.
\newblock Identifying the microtearing modes in the pedestal of diii-d h-modes
  using gyrokinetic simulations.
\newblock {\em Nuclear Fusion}, 2021.

\bibitem{goerler_11}
T.~G\"orler, X.~Lapillonne, S.~Brunner, T.~Dannert, F.~Jenko, F.~Merz, and
  D.~Told.
\newblock The global version of the gyrokinetic turbulence code {GENE}.
\newblock {\em Journal of Computational Physics}, 230(18):7053--7071, August
  2011.

\bibitem{jenko_01b}
F.~Jenko, W.~Dorland, and G.W. Hammett.
\newblock Critical gradient formula for toroidal electron temperature gradient
  modes.
\newblock {\em Phys. Plasmas}, 8:4096, 2001.

\bibitem{pingyu_23}
Ping-Yu Li.
\newblock Etg turbulent transport in mega ampere spherical tokamak (mast)
  pedestal.
\newblock {\em submitted to Nuclear Fusion}.

\bibitem{chapman_23}
B.~Chapman-Oplopoiou and et~al.
\newblock The composition of etg turbulence in jet-ilw pedestals.
\newblock {\em in preparation}.

\bibitem{pueschel_20}
M.J. Pueschel, D.R. Hatch, M.~Kotschenreuther, A.~Ishizawa, and G.~Merlo.
\newblock Multi-scale interactions of microtearing turbulence in the tokamak
  pedestal.
\newblock {\em Nucl. Fusion}, 60(12):124005, December 2020.

\bibitem{belli_23}
E~A Belli, J~Candy, and I~Sfiligoi.
\newblock Spectral transition of multiscale turbulence in the tokamak pedestal.
\newblock {\em Plasma Physics and Controlled Fusion}, 65(2):024001, February
  2023.

\bibitem{field_23}
A.~R. Field, B.~Chapman-Oplopoiou, J.~W. Connor, L.~Frassinetti, D.~R. Hatch,
  C.~M. Roach, S.~Saarelma, and {JET contributors}.
\newblock Comparing pedestal structure in {JET}-{ILW} {H}-mode plasmas with a
  model for stiff {ETG} turbulent heat transport.
\newblock {\em Philosophical Transactions of the Royal Society A: Mathematical,
  Physical and Engineering Sciences}, 381(2242):20210228, February 2023.

\bibitem{frassinetti_17}
L~Frassinetti, S~Saarelma, P~Lomas, I~Nunes, F~Rimini, M~N~A Beurskens,
  P~Bilkova, J~E Boom, E~de~la Luna, E~Delabie, P~Drewelow, J~Flanagan,
  L~Garzotti, C~Giroud, N~Hawks, E~Joffrin, M~Kempenaars, Hyun-Tae Kim,
  U~Kruezi, A~Loarte, B~Lomanowski, I~Lupelli, L~Meneses, C~F Maggi, S~Menmuir,
  M~Peterka, E~Rachlew, M~Romanelli, E~Stefanikova, and JET Contributors.
\newblock Dimensionless scalings of confinement, heat transport and pedestal
  stability in jet-ilw and comparison with jet-c.
\newblock {\em Plasma Physics and Controlled Fusion}, 59(1):014014, 2017.

\bibitem{frassinetti_20}
L.~Frassinetti, S.~Saarelma, G.~Verdoolaege, M.~Groth, J.C. Hillesheim,
  P.~Bilkova, P.~Bohm, M.~Dunne, R.~Fridström, E.~Giovannozzi, F.~Imbeaux,
  B.~Labit, E.~de~la Luna, C.~Maggi, M.~Owsiak, and R.~Scannell.
\newblock Pedestal structure, stability and scalings in {JET}-{ILW}: the
  {EUROfusion} {JET}-{ILW} pedestal database.
\newblock {\em Nuclear Fusion}, 61(1):016001, nov 2020.

\bibitem{frassinetti_21}
L.~Frassinetti, C.~Perez von Thun, B.~Chapman, A.~Fil, J.C. Hillesheim,
  L.~Horvath, G.T.A. Huijsmans, H.~Nyström, V.~Parail, S.~Saarelma,
  G.~Szepesi, B.~Viola, R.~Bianchetti Morales, M.~Dunne, A.R. Field,
  J.~Flanagan, J.M. Fontdecaba, D.~Hatch, B.~Lomanowski, C.F. Maggi,
  S.~Menmuir, S.~Pamela, C.M. Roach, E.~Rachlew, E.R. Solano, and JET
  Contributors.
\newblock Role of the separatrix density in the pedestal performance in
  deuterium low triangularity {JET}-{ILW} plasmas and comparison with {JET}-c.
\newblock {\em Nuclear Fusion}, 61(12):126054, nov 2021.

\bibitem{field_18}
A.~R. Field, L.~Frassinetti, C.~Maggi, S.~Saarelma, and JET contributors.
\newblock Inter-elm power losses and their dependence on pedestal parameters in
  jet-c and iter-like wall h-mode plasmas.
\newblock {\em Proceedings of the 2018 EPS Conference, Prague, Czech Republic,
  2018}, 2018.

\bibitem{kotschenreuther_17}
M.~Kotschenreuther, D.~R. Hatch, S.~Mahajan, P.~Valanju, L.~Zheng, and X.~Liu.
\newblock Pedestal transport in {H}-mode plasmas for fusion gain.
\newblock {\em Nuclear Fusion}, 57(6):064001, 2017.

\bibitem{snyder_09}
P.B. Snyder, R.J. Groebner, A.W. Leonard, T.H. Osborne, and H.R. Wilson.
\newblock Development and validation of a predictive model for the pedestal
  height.
\newblock {\em Phys. Plasmas}, 16:056118, 2009.

\bibitem{larakers_21}
J.~L. Larakers, M.~Curie, D.R. Hatch, R.D. Hazeltine, and S.M. Mahajan.
\newblock Global {Theory} of {Microtearing} {Modes} in the {Tokamak}
  {Pedestal}.
\newblock {\em Physical Review Letters}, 126(22):225001, June 2021.

\bibitem{curie_22}
M.~T. Curie, J.~L. Larakers, D.~R. Hatch, A.~O. Nelson, A.~Diallo, E.~Hassan,
  W.~Guttenfelder, M.~Halfmoon, M.~Kotschenreuther, R.~D. Hazeltine, S.~M.
  Mahajan, R.~J. Groebner, J.~Chen, {DIII-D Team}, C.~Perez~von Thun,
  L.~Frassinetti, S.~Saarelma, C.~Giroud, {JET Contributors}, and M.~M.
  Tennery.
\newblock A survey of pedestal magnetic fluctuations using gyrokinetics and a
  global reduced model for microtearing stability.
\newblock {\em Physics of Plasmas}, 29(4):042503, April 2022.

\bibitem{ashourvan_22}
Arash Ashourvan, R.~Nazikian, and Q.M. Hu.
\newblock Role of the edge stochastic layer in density pump-out by resonant
  magnetic perturbations.
\newblock {\em Nuclear Fusion}, 62(7):076007, apr 2022.

\bibitem{verdoolaege_21}
G.~Verdoolaege, S.M. Kaye, C.~Angioni, O.J.W.F. Kardaun, M.~Maslov,
  M.~Romanelli, F.~Ryter, K.~Thomsen, the ASDEX Upgrade~Team, the EUROfusion
  MST1~Team, and Jet Contributors.
\newblock The updated {ITPA} global {H}-mode confinement database: description
  and analysis.
\newblock {\em Nuclear Fusion}, 61(7):076006, July 2021.

\bibitem{lomanowski_22}
B.~Lomanowski, M.~Dunne, N.~Vianello, S.~Aleiferis, M.~Brix, J.~Canik, I.S.
  Carvalho, L.~Frassinetti, D.~Frigione, L.~Garzotti, M.~Groth, A.~Meigs,
  S.~Menmuir, M.~Maslov, T.~Pereira, C.~Perez~von Thun, M.~Reinke, D.~Refy,
  F.~Rimini, G.~Rubino, P.A. Schneider, G.~Sergienko, A.~Uccello,
  D.~Van~Eester, and {JET Contributors}.
\newblock Experimental study on the role of the target electron temperature as
  a key parameter linking recycling to plasma performance in {JET}-{ILW}*.
\newblock {\em Nuclear Fusion}, 62(6):066030, June 2022.

\bibitem{bourdelle_23}
Clarisse Bourdelle, J~Morales, Jean-Francois Artaud, Ondrej Grover, Tennessee
  Radenac, Jerome~B Bucalossi, Yann Camenen, Guido Ciraolo, Frederic Clairet,
  Remi Dumont, Nicolas Fedorczak, Jonathan Gaspar, Christophe Gil, Marc
  Goniche, Christophe Guillemaut, James~Paul Gunn, Patrick Maget, Pierre Manas,
  Valeria Ostuni, Bernard Pegourie, Yves Peysson, Patrick Tamain, L~Vermare,
  and Didier Vezinet.
\newblock Separatrix parameters and core performances across {WEST} {L}-mode
  database.
\newblock {\em Nuclear Fusion}, February 2023.

\end{thebibliography}
\bibliographystyle{unsrt}

\appendix

\cleardoublepage

\end{document}